\documentclass[mathleft]{an}
\usepackage{graphicx}
\usepackage{times}
\newcommand{\Mearth}{$M_\oplus$}
\newcommand{\Msun}{$M_\odot$}
\newcommand{\Mjup}{$M_\mathrm{J}$}

\newcommand{\AU}{{\sc au}}

\newcommand{\corot}{\emph{CoRoT}}
\newcommand{\kepler}{\emph{Kepler}}

\begin{document}

\Pagespan{789}{}
\Yearpublication{2006}%
\Yearsubmission{2005}%
\Month{11}%
\Volume{999}%
\Issue{88}%

\title{Are extrasolar oceans common throughout the Galaxy?}

\author{%
    David Ehrenreich\inst{1}\fnmsep\thanks{Corresponding author:
        \email{ehrenreich@iap.fr}\newline}
    \and
    Arnaud Cassan\inst{2,3}}

\titlerunning{Are oceans common throughout the Galaxy?}
\authorrunning{D. Ehrenreich \& A. Cassan}

\institute{%
    Institut d'astrophysique de Paris, CNRS (UMR 7095), Universit\'e
        Pierre et Marie Curie, 98 bis boulevard Arago, 75014 Paris, France
    \and
    Astronomisches Rechen-Institut, Zentrum f\"ur Astronomie der Universit\"at
        Heidelberg, M\"onchhofstrasse 12-14, 69120 Heidelberg, Germany
    \and
    The PLANET collaboration}

\received{} \accepted{} \publonline{later}

\keywords{planets and satellites: general --- planets and satellites:
individual (OGLE 2005-BLG-390Lb) --- techniques: microlensing}

\abstract{Light and cold extrasolar planets such as OGLE 2005-BLG-390Lb, a 5.5
Earth-mass planet detected via microlensing, could be frequent in the Galaxy
according to some preliminary results from microlensing experiments. These
planets can be frozen rocky- or ocean-planets, situated beyond the snow line
and, therefore, beyond the habitable zone of their system. They can nonetheless
host a layer of liquid water, heated by radiogenic energy, underneath an ice
shell surface for billions of years, before freezing completely. These results
suggest that oceans under ice, like those suspected to be present on icy moons
in the Solar system, could be a common feature of cold low-mass extrasolar
planets.}

\maketitle

\section{Introduction}
More than 200 extrasolar planets have been detected during the past 15 years.
Aside from the planetary search by itself, the refinement and multiplication of
detection techniques have allowed to determine several physical properties of
the targeted planets, including their mean density, hence their nature. Roughly
90\% of the detected planets\footnote{See the \emph{Extrasolar planets
encyclopaedia} (http://exoplanet.eu).} have masses within two orders of
magnitude of Jupiter's mass (\Mjup); they are thought to be gaseous giants.
Combined planetary mass and radius measurements have confirmed this fact for
nearly all planets transiting their stars, for which a radius measurement is
possible.

Transiting planets less massive than `hot Jupiters' will hopefully be found by
transit search missions \corot\ (flying) and \kepler\ (launch scheduled for
2008). Meanwhile, it is rather difficult to infer the natures, either gaseous,
icy, or rocky, of Uranus- and lower-mass planets ($\la 15$ Earth masses,
\Mearth). In fact, the `critical' core mass usually considered to separate the
formation processes of telluric and giant planets in the Solar system is $\sim
8$~\Mearth\ (Wuchterl et al.\ 2000). This theoretical limit, however, should
not be applied \emph{as it is} to extrasolar planets, as illustrated by the
ambiguous case of HD\,149026b, a $\sim$~\Mjup\ planet suspected to host a
massive dense core in order to explain its small radius (Sato et al.\ 2005;
Fortney et al.\ 2006; Broeg \& Wuchterl 2007).

Foreseeing the possible diversity of low-mass planets, models of different
internal structures and atmospheres have been recently flourishing. These
speculative descriptions basically distinguish between Earth-like rocky planets
(Valencia et al.\ 2006, 2007; Sotin et al.\ 2007) and more water-rich planets,
hypothetical `ocean-planets' (Kuchner 2003; L\'eger et al.\ 2004; Sotin et al.\
2007; Selsis et al.\ 2007). The detections of planets GJ\,876d (7.5~\Mearth,
Rivera et al.\ 2005) by radial velocimetry and OGLE 2005-BLG-390Lb
(5.5~\Mearth, Beaulieu et al.\ 2006) by microlensing justify these modeling
approaches and allow the first model applications.

Ehrenreich et al.\ (2006; hereafter, E06) have used observational constrains
from Beaulieu et al.\ (2006) together with modeling from Sotin et al.\ (2007)
and Hussmann et al.\ (2002) to characterize OGLE 2005-BLG-390Lb. They raised
the possibility that this cold ($\sim40$~K) low-mass planet could host a liquid
water ocean under an ice shell surface, similarly to some icy moons in the
Solar System, according to current models (see, e.g., Spohn \& Schubert 2003;
Hussmann et al.\ 2006).

In this speculative paper, we combine the first results on microlensing
searches detection efficiency and the results of E06, to highlight the fact
that planets similar to OGLE 2005-BLG-390Lb could be more abundant than gaseous
giants---and potentially common throughout the Galaxy. This would make them
particularly interesting objects to study, especially if the conditions
allowing for the existence of oceans are fulfilled.

\section{Microlensing detection efficiency}
Current ground-based microlensing searches probe distant (at several kpc) cool
planetary companions in the range 1--10~astronomical units (\AU),
preferentially around M- and K-dwarf hosts, with masses down to a few Earths.
The PLANET/Robonet (Probing Lensing Anomalies NETwork) collaboration operates a
round-the-clock follow-up on ongoing microlensing events towards the Galactic
bulge, by means of currently eight 1--2m-class telescopes situated in the south
hemisphere.

Some promising detections have recently been made, with the discoveries of two
gas giants of a few Jupiter masses, MOA 2003-BLG-53Lb (Bond et al.\ 2004) and
OGLE 2005-BLG-071Lb (Udalski et al.\ 2005), a Neptune-mass planet OGLE
2005-BLG-169Lb (Gould et al.\ 2006) and the rocky/icy planet OGLE
2005-BLG-390Lb (Beaulieu et al.\ 2006).

Kubas et al.\ (2007, submitted) have computed the detection probability for
OGLE 2005-BLG-390Lb-like planets, and find that, assuming a perfect
observational setup, it does not exceed $3\%$. Furthermore, they also compute
the probability of detecting an additional Jupiter-mass planet to the system,
and find it is greater than $50\%$ for orbits between $1.1$ and $2.3$~\AU.
Therefore, OGLE 2005-BLG-390Lb was detected in spite of a low detection
probability, while no giant companion was found, even with a rather high
detection efficiency. This fact supports the core accretion theory which
predicts that sub-Neptune mass planets are more common than giants around M
dwarfs, as also pointed out by Gould et al.\ (2006).

One can go further in the conclusion by computing the detection efficiency of
the microlensing searches. This is done by determining for many microlensing
events which do not show obvious planetary signature, to which extent the data
rule out the presence of a planetary companion to the lens. While a detailed
and complete analysis including 11 years of PLANET observations (1995--2005)
will be provided by Cassan et al.\ (2007, in prep.), a preliminary result can
be derived from the 2004 season (Cassan \& Kubas 2007).

\begin{figure}
\resizebox{\hsize}{!}{\includegraphics{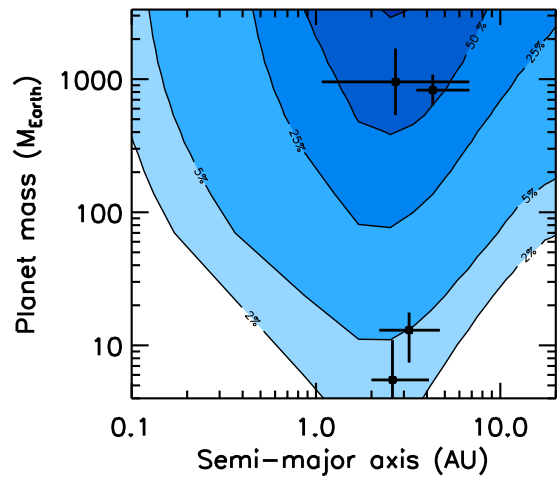}} \caption{Average detection
efficiency of planets as a function of mass and orbital separation (assuming
circular orbits) in the 14 favourable events monitored by PLANET in 2004
(preliminary analysis) along with the planets detected by microlensing as of
2006 marked as dots. For Jupiter-mass planets, the detection efficiency reaches
$50\%$, while it decreases only with the square-root of the planet mass until
the detection of planets is further suppressed by the finite size of the source
stars for planets with a few Earth masses. Nevertheless, the detection
efficiency still remains a few per cent for rocky/icy planets below 10 Earth
masses.} \label{fig:eff}
\end{figure}

Figure~\ref{fig:eff} presents the mean detection efficiency iso-contours (in
percent) as a function of planet mass and separation, combining 14 well sampled
events from the 2004 PLANET season alone. This diagram is produced by computing
for every individual event their detection efficiency as a function of the
planet-to-star mass ratio and the instantaneous separation, and use an
appropriate Galactic model with a Bayesian analysis (Dominik 2006) to convert
the parameters in physical units of the planet properties (mass and separation
in \AU). The detected planets are over-plotted on the same figure, which allows
to give preliminary qualitative conclusion.

Hence, combining detections and detection efficiency, one can evaluate which
kind of planets can be `easily' found but are actually not, and those which
were detected despite a small probability. It appears that in spite of the
difference in detection efficiencies for $\sim1$~\Mjup\ and 1--15~\Mearth\
planets, the same number of objects has been evidenced in each category. This
suggests that 1--15~\Mearth\ planets are more common around low-mass stars than
giant gaseous planets.

Of course, we shall temperate this affirmation given the small number of
detections so far; however, it is in accordance with the core-accretion theory
of planet formation around low-mass stars (Laughlin et al.\ 2004). Since those
stars represent the major fraction of the stars in the Galactic disk (Chabrier
2001), the observations of Gould et al.\ (2006) and Beaulieu et al.\ (2006)
implies that low-mass planets are frequent in the Milky Way.

\section{Oceans in frozen low-mass planets}
According to Beaulieu et al.\ (2006) calculated uncertainties, OGLE
2005-BLG-390Lb has a mass between 3 and 11 \Mearth\ and a semi-major axis
between 2 and 4~\AU. The mass of the parent star is between 0.1 and 0.4~solar
mass (\Msun). It is potentially the lightest extrasolar planet detected so far,
and according to the previous section and the rather large errorbars associated
with its mass and semi-major axis, this planet could be representative of a
large fraction of planets around M dwarfs. Such stars would provide very few
energy per surface unit to the distant planet (2--4~\AU), typically that of
Pluto in the Solar system ($\sim0.1$~W\,m$^{-2}$). Hence, the planet can be
either a cold and massive analog of the Earth, or a frozen ocean-planet
(L\'eger et al.\ 2004). In both cases, it could have a solid icy surface with
temperatures of $\sim40$~K. In spite of that, E06 showed that it could host a
subsurface liquid water ocean under the ice shell surface, depending on the
ice-to-rock mass ratio and age of the planet/star.

\subsection{The ice-to-rock mass ratio}
The total mass of ice in the planet relatively to the quantity of refractory
elements (metals and rocks), the ice-to-rock mass ratio ($I/R$), is treated as
a free parameter in E06. They used the mass-radius relation from Sotin et al.\
(2007) to determine the planetary radius for different $I/R$, namely
$\sim10^{-4}$, 0.33, and~1. Since ice is less dense than rock, a planet
containing more ice is also larger than a same-mass rocky planet.

The first case, $I/R \sim 10^{-4}$, in fact applies to a rocky planet with a
H$_2$O content similar to the Earth one; there, an ice layer covering the rocky
mantle would be rather thin ($\la10$~km) compared to the planetary radius, and
composed exclusively of low-pressure ice. In the last two cases, $I/R \sim
0.33$ and $I/R \sim 1$, the planet could be a massive intermediate between
Europa and Ganymede, and a hypothetical ocean-planet, respectively. There, a
non-negligible fraction of the planetary mass is accounted for by water, and
the thick ice shell must be largely composed of high-pressure ice due to the
strong gravity. However, E06 distinguish between a thin ($\la60$~km)
low-pressure ice overtopping a thick ($\sim1\,000$~km) high-pressure icy
mantle. The negative slope of the low-pressure H$_2$O ice melting curve as a
function of pressure makes the existence of a $\la50$-km-thick ocean possible
at the bottom of the low-pressure ice layer. The different possibilities are
sketched in Fig.~\ref{fig:struct}.

\subsection{Heating \emph{versus} cooling of the subsurface ocean}
\label{sec:heatflow} Knowing that the age of the planet and star is
$\sim10$~Gyr, a typical age for a star of the Bulge, and given the negligible
irradiation the planet is receiving at its surface ($\sim0.1$~W\,m$^{-2}$), the
existence of liquid water requires an internal heat source capable of melting
the ice in the depths. This source could be the decay of radioactive long-lived
isotopes of uranium, thorium, and potassium, possibly included in the rocky
mantle. The internal heat production per unit of mass is $h = f \sum_i h_i
X_i(t)$, where $h_i$ is the heat produced by the isotope $i$ per unit of mass,
$X_i(t)$ is the mass fraction of this isotope at a time $t$, and $f$ is the
mass fraction of rocks in the planet. The more rocks within the planet---i.e.,
the lower the $I/R$ ratio---and the younger it is, the larger the heat
production.

To have a liquid layer below the ice shell, the temperature must reach at least
$\sim250$~K above the point where the slope of the melting curve becomes
positive with pressure/depth. This point is situated at a pressure of
$\sim0.2$~GPa, which corresponds to a depth $<100$~km for the range of
planetary masses and radii considered here. In other words, the top of the
ocean cannot be deeper than 100~km beneath the surface.

This thin ice shell surface, relatively to the size of the planet, can be even
thinner depending on the total quantity of ice in the planet. For $I/R \sim
10^{-4}$, the ice shell cannot be thicker than $\sim10$~km. In this case, E06
showed that the heat loss is too fast to allow the survival of a subsurface
ocean after 10~Gyr---the age supposed for OGLE 2005-BLG-390Lb---despite a large
heat production rate. The maximum 100-km thickness ice shell better isolating
the liquid layer from the cold surface is reached in models with $I/R \sim 1$
(ocean-planets). However, as we seen above, the heat production is too low to
make the temperature reach the critical 250~K at this depth.

In fact, a `compromise' value could be found around $I/R \sim 0.33$, in order
to have both a sufficient heat production rate and a sufficiently thick ice
shell surface. Therefore, icy planets with a composition between that of Europa
($I/R \sim 0.1$) and Ganymede ($I/R \sim 0.5$), may be better place to find an
ocean under an ice shell.

\begin{figure}
\resizebox{\hsize}{!}{\includegraphics{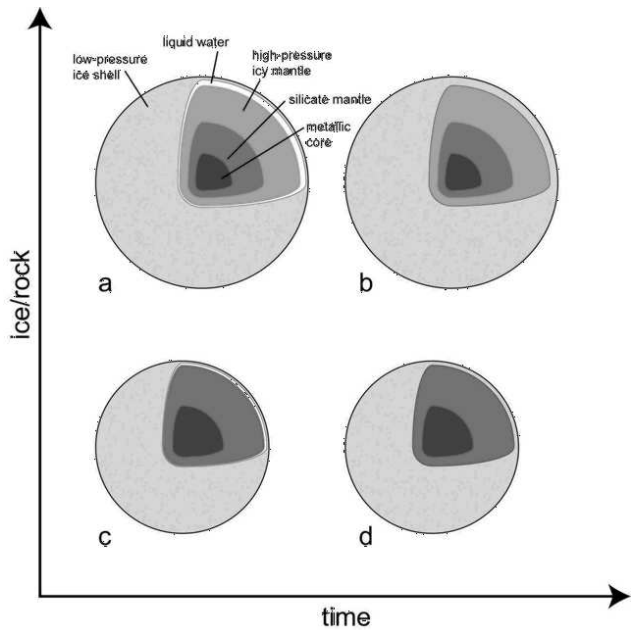}} \caption{Sketches of the
possible internal structures of OGLE 2005-BLG-390Lb and analog planets in the
same range of mass/semi-major axis. Panel $a$ shows a young planet with $I/R$
typically $\sim1$, where an ocean (\emph{white}) is found beneath a thin ice
shell surface. The planet is composed by a metallic core, rocky mantle, and
high-pressure icy mantle (\emph{from darker to lighter grey}). Panel $b$ shows
the same planet at a more advanced age, where the radioactive heating has
decreased and the ocean has condensed. The same evolution is figured for a
rocky planet ($I/R \sim 10^{-4}$) in panels $c$ and $d$. In this case, the
planet is mostly composed by a metallic core and a rocky mantle, overtopped by
a tenuous ice shell. See E06 and Sotin et al.\ (2007) for details.}
\label{fig:struct}
\end{figure}

In any cases, E06 concluded that owing to its age ($\sim10$~Gyr), OGLE
2005-BLG-390Lb is most likely entirely frozen (Figs.~\ref{fig:struct}$b$
or~\ref{fig:struct}$d$). It is however possible to roughly estimate since how
long the ocean is condensed. The present value of the heat production should be
increased by a factor 2.1 to 2.5 in order to have liquid water in the
10-Gyr-old planet, assuming $I/R \sim 0.33$. These values of the heat
production rate were in fact reached when the planet was $\sim4$-Gyr old.

This estimation assumes that radiogenic heating is the only heat source in the
planet interior. However, it should be regarded as a minimum value since it
does not consider the remnant heat from the accretion, which is certainly not
negligible, in particular for such a massive planet. This allows to suggest
that liquid water was certainly present in the past during several billion
years below the surface of OGLE 2005-BLG-390Lb, and might be commonly found on
similar but younger planets.

\section{Conclusions}
First results from microlensing surveys suggest that planets less massive than
Uranus or Neptune are more common around low-mass stars than giant gaseous
planets, as proposed by the core-accretion theory. Low-mass stars, in turn, are
the most represented stellar population in the Galaxy. It is therefore relevant
to study what kind of planets they could host. They could be similar to OGLE
2005-BLG-390Lb, a few times more massive than Earth and completely frozen,
unless they are composed by about 3/4 of rocks and metals and 1/4 of ice, in
mass. This `recipe' indeed favors the presence of a liquid layer beneath a
frozen surface. It is close to the composition of icy satellites in the Solar
system, such as Europa or Ganymede. These objects have long been suspected of
hosting liquid water---a suspicion that has now spread onto less massive
satellites and Kuiper belt objects (Hussmann et al.\ 2006), and on low-mass
extrasolar planets (L\'eger et al.\ 2004; E06).

It implies that `oceans under ice' could be a common feature of low-mass
extrasolar planets, largely spread around low-mass stars in the Galaxy. This is
an interesting perspective for astrobiology, because oceans, even under an ice
shell, are \emph{a priori} good places for the apparition/subsistence of life.

In return, this underlines the needs and highlights the interests for studying
Solar system frozen bodies, such as icy moons and Kuiper belt objects,
especially through space exploration missions capable of \emph{in situ} probing
of the oceans potentially present there. These oceans might be common
throughout the Galaxy.

\acknowledgements We thank Jean-Philippe Beaulieu for rampant ideas and useful
insights, the organizers in Budapest for a scientifically and socially
enriching meeting, and the British Council for making the meeting possible.

\newpage

\end{document}